\documentclass[10pt]{article}
\usepackage{authblk}

\usepackage{amssymb}
\usepackage{amsthm}
\usepackage{xcolor}

\newcommand{\TNS}{Ta$_2$NiSe$_5$}
\newcommand{\EF}{$E_{\text F}$}
\newcommand{\TC}{$T_{\text C}$}
\newcommand{\SM}{\textcolor{black}}

\usepackage{hyperref}
\usepackage{caption}
\newcommand{\Fref}{Fig.~\ref}

\usepackage{amsmath}
\usepackage{soul}
\usepackage{nicefrac}
\usepackage{ulem}
\usepackage{graphicx}

\begin{document}

%\begin{frontmatter}
\title{Ultrafast charge carrier and exciton dynamics in an excitonic insulator probed by time-resolved photoemission spectroscopy}
\author[1,2]{Selene Mor\thanks{selene.mor@unicatt.it}}%\corref{cor}} %
%\cortext[cor]{selene.mor@unicatt.it}
%\affiliation[inst2]{organization={Department of Physics and I-LAMP Research Center, Universit\'{a} Cattolica del Sacro Cuore},
   %         addressline={via della Garzetta 48}, 
      %      city={Brescia},
         %   postcode={25133},
            %country={Italy}}
\affil[1]{Department of Physical Chemistry, Fritz Haber Institute of the Max Planck Society, Faradayweg 9, Berlin, 14195, Germany}
\affil[2]{Department of Physics and I-LAMP Research Center, Universit\'{a} Cattolica del Sacro Cuore, via della Garzetta 48, Brescia, 25133, Italy}
            
%\affiliation[inst1]{organization={Department of Physical Chemistry, Fritz Haber Institute of the Max Planck Society},
   %         addressline={Faradayweg 9}, 
      %      city={Berlin},
         %   postcode={14195},
            %country={Germany}}

\author[3,1]{Marc Herzog}

\author[5]{Claude Monney}

\author[4,1]{Julia St\"{a}hler}

%\affiliation[inst3]{organization={Institute of Physics and Astronomy,  University of Potsdam},
   %         addressline={Karl-Liebknecht-Straße 24/25}, 
      %      city={Potsdam-Golm},
         %   postcode={14476},
            %country={Germany}}
\affil[3]{Institute of Physics and Astronomy,  University of Potsdam, Karl-Liebknecht-Straße 24/25, Potsdam-Golm, 14476, Germany}
           
%\affiliation[inst5]{organization={Department of Physics and Fribourg Center for Nanomaterials, University of Fribourg},
   %         addressline={Ch. du Mus\'{e}e 3}, 
      %      city={Fribourg},
         %   postcode={1700},
            %country={Switzerland}}
\affil[5]{Department of Physics and Fribourg Center for Nanomaterials, University of Fribourg, Ch. du Mus\'{e}e 3, Fribourg, 1700, Switzerland}

%\affiliation[inst4]{organization={Department of Chemistry, Humboldt-Universit\"{a}t zu Berlin},
   %         addressline={Brook-Taylor-Stra\ss e}, 
      %      city={Berlin},
         %   postcode={12489},
            %country={Germany}}
\affil[4]{Department of Chemistry, Humboldt-Universit\"{a}t zu Berlin, Brook-Taylor-Stra\ss e, Berlin, 12489, Germany}

\maketitle
\begin{abstract}
An excitonic insulator phase is expected to arise from the spontaneous formation of electron–hole pairs (excitons) in semiconductors  where the exciton binding energy exceeds the size of the electronic band gap. At low temperature, these ground state excitons stabilize a new phase by condensing at lower energy than the electrons at the valence band top, thereby widening the electronic band gap. The envisioned opportunity to explore many-boson phenomena in an excitonic insulator system is triggering a very active debate on how ground state excitons can be experimentally evidenced. Here, we employ a nonequilibrium approach to spectrally disentangle the photoinduced dynamics of an exciton condensate from the entwined signature of the valence band electrons. By means of time- and angle-resolved photoemission spectroscopy of the occupied and unoccupied electronic states, we follow the complementary dynamics of conduction and valence band electrons in \SM{the photoexcited low-temperature phase of }\TNS, the hitherto most promising single-crystal candidate to undergo a semiconductor-to-excitonic-insulator phase transition. The photoexcited conduction electrons are found to relax within less than 1~ps. Their relaxation time is inversely proportional to their excess energy, a dependence that we attribute to the reduced screening of Coulomb interaction and the low dimensionality of \TNS. Long after ($>$~10~ps) the conduction band has emptied, the photoemission intensity below the Fermi energy has not fully recovered the equilibrium value. Notably, this seeming carrier imbalance cannot be rationalized simply by the relaxation of photoexcited electrons and holes across the semiconducting band gap. Rather, a rate equation model involving different photoemission crosssections of the valence electrons and the condensed excitons is able to reproduce the delayed recovery of the photoemission intensity below the Fermi energy. The model shows that electron quantum tunnelling between the exciton condensate and the valence band top is enabled by an extremely small activation energy of $4 \times 10^{\text{-6}}$~eV and explains the retarded recovery of the exciton condensate. Our findings not only determine the energy gain of ground state exciton formation with exceptional energy resolution, but also demonstrate the use of time-resolved photoemission to unveil the re-formation dynamics of an exciton condensate with femtosecond time resolution. 
\end{abstract}

%\begin{keyword}
%Ultrafast quasiparticle relaxation dynamics \sep Excitonic insulator \sep Ground-state exciton formation and dissociation \sep Time- and angle-resolved photoemission spectroscopy  \sep Two-photon photoelectron spectroscopy
%\end{keyword}

%\end{frontmatter}

\tableofcontents
%% main text
\section{Introduction}
\label{sec:intro}
The excitonic insulator (EI) phase was theoretically proposed in the 1960s as an insulating phase that arises from the spontaneous formation of bound electron-hole pairs (excitons) at equilibrium~\cite{Mott1961,Jero1967,Kohn1967}. During a transition to the EI phase, a weak screening of the electron-hole Coulomb interaction leads to an exciton binding energy exceeding %\SM{\sout{, possibly even slightly,}}
the fundamental electronic band gap. This condition favors the formation of excitons from a fraction of valence electrons that thereby lower their energy and contribute, at sufficiently low temperatures, to a condensate of bosonic quasiparticles~\cite{Kohn1967,Bron2006,Seki11, Zenker12,Halperin68,DiBart81}. 
\begin{figure}
\centering
\includegraphics[width=0.3\textwidth]{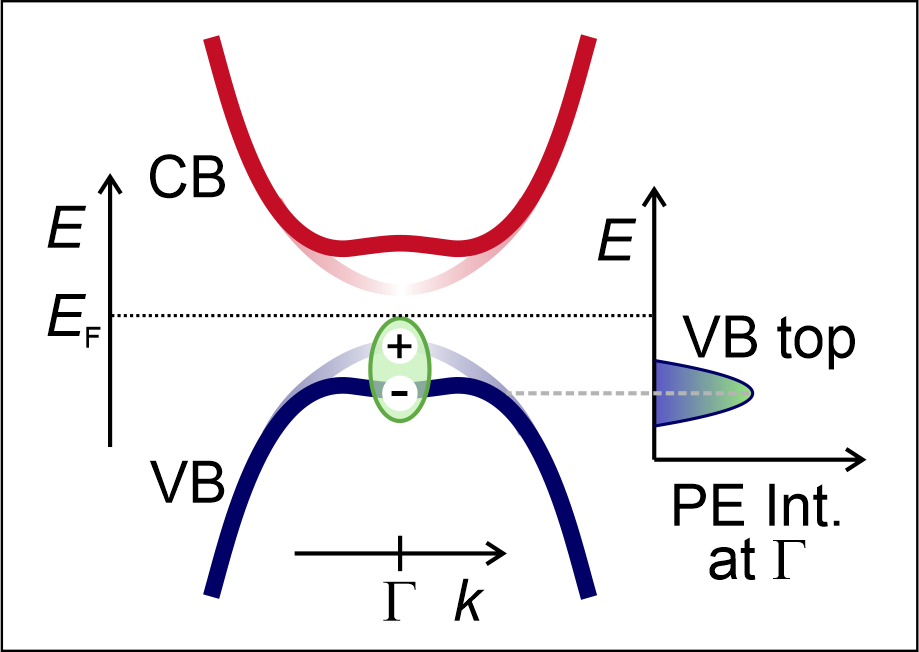}
		\caption[]{Sketch of the electronic band structure modified by the ground state exciton condensate. \SM{In the EI phase, the PE signal arising from the excitonic state (represented by the +/- bound state) merges with the PE signal from the renormalized flat VB top of the single particle band structure. The relative energies of single and many-body electronic states are aligned as they are measured in the PE experiment.} 
 }
	\label{EIARPES}
\end{figure}
The resulting \SM{modification of the single-particle} electronic band structure is depicted in \Fref{EIARPES}, where an electron at the valence band (VB) top is pushed towards a higher binding energy after spontaneously binding with a hole. The downwards energy shift represents the energy gain in forming a ground state exciton.% In order to stabilize the new EI ground state, a lattice instability is expected to evolve hand in hand with the electronic phase transition.

The interest in corroborating the theory predictions and the premise of providing a material platform for the study of quantum many-boson phenomena have driven several experimental efforts in the past few decades aiming at finding a real system which hosts an EI condensate in the ground state. A plethora of compounds have been investigated: double layers of GaAs~\cite{Eisenstein92,Tutuc04,Nandi12}, AlAs/GaAs~\cite{Butov94} and graphene~\cite{Gorbachev12,Li17, Liu17} to measure collective exciton transport driven by quantum-Hall voltage; two-dimensional heterobilayers of transition metal dichalcogenides semiconductors~\cite{Fogler14,Wang19,Sigl20, Ma21} to detect interlayer tunneling current enabled by electron-hole pairing across the two layers. Angle-resolved photoemission spectroscopy (ARPES) of the semimetal TiSe$_2$ has revealed an anomalous dispersion at the VB top below $T$~=~250~K~\cite{Traum78,Pillo00,Rossnagel02,Cerc2007,Lian20}. This spectral feature has been explained as a result of the coupling of the VB and the conduction band (CB) at low temperatures, leading to spontaneous exciton formation. Furthermore, momentum-resolved electron energy-loss spectroscopy of TiSe$_2$ at low temperatures~\cite{Kogar17} assigned the energy collapse of a collective electronic mode to the occurrence of exciton condensation. Yet, the concomitant formation of a charge density wave and a periodic lattice distortion has left the question unanswered on whether the opening of the excitonic gap at low temperatures is driven by a mechanism of pure electronic origin.

The direct semiconductor \TNS, subject of the present work, is attracting enormous attention due to its semiconductor-to-insulator electronic phase transition combined with an orthorhombic-to-monoclinic structural symmetry change at the critical temperature \TC~=~328~K~\cite{DiSa1986,Sunshine1985,Seo18}. ARPES studies below \TC~discovered a flattening of the VB top at $\Gamma$, i.e. at $k$~=~0, reminiscent of what was observed in TiSe$_2$. In combination with finite-temperature simulations, such feature has been assigned to the spontaneous coupling of the VB and CB and the realization of an EI ground state~\cite{Waki2009,Waki2012,Kane2015, Seki2014}. At the same \TC, the crystal symmetry of \TNS~changes~\cite{Kane2013,Lu17}, yet without involving periodic lattice distortion nor producing a charge density wave. Even so, recent theoretical~\cite{Mazza2020,Windgaetter2021} and experimental~\cite{Watson2020} studies pointed out that the lattice distortion coupled to the electronic instability is in fact needed to open the band gap. Finally, a recent ARPES study of the VB top of \TNS~above \TC~aimed at the extraction of the entwined photoemission (PE) signature of a precursor state of spontaneously formed excitons~\cite{Fukutani21}. 

All the discussed observations have being setting \TNS~as an ideal candidate to investigate the EI formation mechanism. Towards this goal, ultrafast spectroscopy techniques have further proven to offer a powerful probe. The approach is based on the optical excitation of the system's ground state followed by the determination of the relaxation pathways of the electronic, excitonic and structural degrees of freedom on the relevant time and energy scales~\cite{Wegk15,Giannetti16,Smallwood16,Maiuri20,Lloyd-Hughes21, Mor2021}. \SM{It is important to note that the screening of the Coulomb interaction changes upon photoexcitation of quasiparticles in the system. More precisely, carriers and excitons enhance the screening of the Coulomb interaction, which has, beyond others~\cite{Wang20}, two important consequences: (i) Stronger screening steepens the attractive Coulomb potential of the holes in real space, leading to reduced binding energies of the excitons~\cite{Silkin2015,Cui2014}, and (ii) the Hartree contribution to the self energy is modified, causing band gap renormalization~\cite{Chernikov15, Pogna16, Cunningham17,Sie17,Calati21,Calati22}. In the case of an excitonic insulator, both effects lead to a global upward shift of the VB-top PE signature, which involves electronic states of the single particle band structure as well as electrons in the exciton condensate. The upward shift of the VB or redshift of the absorption has been consistently observed in recent PE and optical measurements~\cite{Por14,Mor2017,Okazaki2018,Mitsuoka20,Saha2021}.} In \TNS, time-resolved optical spectroscopy studies below \TC~were able to retrieve coherent oscillations, quenching and recovery of collective excitonic orders~\cite{Werdehausen2018,Bretscher2021,Bretscher21bis}. Our coherent optical phonon spectroscopy study discovered that the low-temperature monoclinic symmetry persists against a photoinduced structural phase transition~\cite{Mor2018} as a result of optical absorption saturation occurring at fluences higher than 0.2~mJ/cm$^2$~\cite{Mor2017}. \SM{%\sout{To some extend similarly,}
Similarly,} broadband ultrafast optical spectroscopy works proved the absence of a structural phase transition for fluences up to 1~mJ/cm$^2$ at 10~K~\cite{Miyamoto22} and 0.5-0.6~mJ/cm$^2$ even at 300~K~\cite{Bretscher2021}. Moreover, a recent study reported, upon photoexcitation beyond few mJ/cm$^2$, a steady state characterized by a lattice structure which is different from any of the thermally accessible equilibrium ones~\cite{Liu21}. In our time- and angle-resolved photoemission spectroscopy (tr-ARPES) work combined with Hartree-Fock calculations~\cite{Mor2017}, we demonstrated that a transient enhancement of the electronic band gap of \TNS~is enabled by strong electron-hole correlations when the system remains out of equilibrium. This EI band gap enhancement becomes sizable upon excitation above the optical absorption saturation threshold and dominates over the renormalization of the fundamental semiconductor band gap due to free-carrier-induced screening of the Coulomb interaction which has been reported by various other tr-ARPES experiments exploring much higher excitation density ranges above 1~mJ~cm$^{-1}$ \cite{Okazaki2018,Mitsuoka20,Baldini20,Saha2021}. 

%All-in-all, rich knowledge has been gained on the nonequilibrium properties of \TNS\, and the impact of collective excitations (excitons and phonons) on its ultrafast dynamics. 
All-in-all, rich knowledge has been gained on the impact of collective excitations (excitons and phonons) on the nonequilibrium behavior of \TNS\, below \TC. However, a quantitative discernment of the exciton condensate (XC) from the electron population at the VB top has not been established in PE spectroscopy. Particularly, the dynamics of the VB electrons that do not participate in the XC has remained an entangled information in the recorded spectra. This originates from the entwined nature of ground state excitons and valence electrons, and the expected energetic vicinity of their spectroscopic signatures, as depicted in \Fref{EIARPES}. Spectral and temporal disentanglement of the two features would represent a remarkable step forward in addressing the EI phase. 

In the present work, we employ tr-ARPES (see Section~\ref{sec:exper}) of both the occupied and unoccupied electronic band structure of \TNS\, to unravel the composite relaxation pathway of an electron excited to the CB back to the EI ground state. Our approach allows us to quantify extremely small energy differences that govern the formation of ground state excitons and determine the ultrafast dynamics of the XC in \TNS.

By combining one- and two-photon photoelectron spectroscopy, we initially obtain momentum-resolved mapping of the occupied and unoccupied electronic band structure of the low-temperature phase of \TNS, and compare it with density-functional-theory (DFT) calculations~\cite{Monney2020}~(see Section~\ref{sec:sec3}). \SM{We then look at the photoexcited low-temperature EI phase \textit{without} driving the transition.} Particularly, the ultrafast dynamics of the excited states is addressed by tr-ARPES of the CBs~(see Section~\ref{sec:sec4}). First, we corroborate our previously reported observation of an optical absorption saturation threshold~\cite{Mor2017} by showing that the electron population resonantly excited to high-energy states at $\Gamma$ also saturates at comparable fluence values. Then, we connect the energy- and fluence-dependent relaxation dynamics of the CB electrons to \SM{%\sout{the quasi-one dimensionality} 
scattering events within a band involving two-dimensional states with linear dispersion and to the reduced screening of the Coulomb interaction of \TNS.} 
Unexpectedly, we find that the PE intensity of the CB minimum (CBM) decays much faster ($<$~2~ps) than the recovery of the VB-top PE intensity ($>$~10~ps), as if the decay of the VB-top hole population would not evolve along with the CBM depopulation. From this observation, we construct a rate equation model to fit the transient PE intensity of the VB top and achieve the discrimination of the photoinduced dynamics of ground state excitons from those of the valence electrons at $\Gamma$~(see Section~\ref{sec:sec5}). The model includes the CB electrons decay across the band gap, the VB top re-filling by electrons, and the re-formation of ground state excitons in the condensate. We find that the bottleneck for the recovery of the equilibrium PE intensity of the VB top is due to the annihilation of ground state excitons which delays the restoration of the equilibrium XC population \SM{particularly when the photoexcitation density is increased}. Our study not only clarifies how transient \SM{screening changes of the} Coulomb interaction \SM{by photoexcited free carriers} and the re-formation and annihilation of ground state excitons impact on the quasiparticle relaxation dynamics of \TNS, but also show the strength of tr-ARPES in detecting the signature of an XC in the ultrafast time domain.

\section{Methods}
\label{sec:exper}
\subsection{\texorpdfstring{Ta\textsubscript{2}NiSe\textsubscript{5}}~ sample preparation}
\begin{figure}
\centering
\includegraphics[]{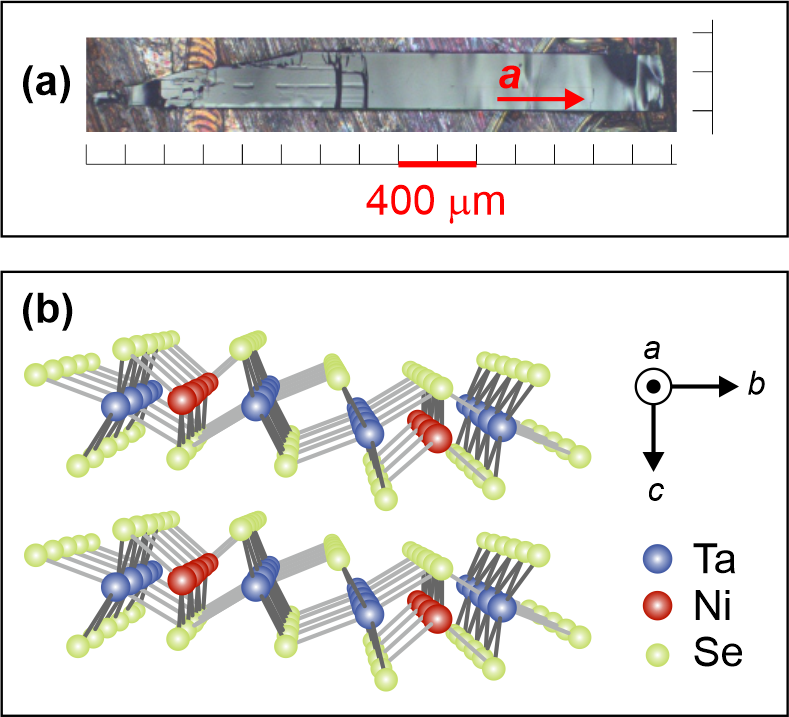}
		\caption[]{(a) Cleaved surface of \TNS. (b) Crystal structure of \TNS. Adapted from~\cite{Thesis}.}
	\label{Fig0}
\end{figure}
We investigate \TNS\,single crystals grown by chemical vapor transport~\cite{Sunshine1985}, with lateral size of 0.5~$\times$~10~mm$^2$ and thickness of 0.4~mm (see \Fref{Fig0}(a)). The longest sample edge corresponds to the in-plane crystallographic $a$ axis parallel to the one-dimensional Ta and Ni atomic chains (see \Fref{Fig0}(b)). The thickness results from the stacking of van-der-Waals bonded $a$-$b$ atomic planes along the crystallographic $c$ axis. 

Prior to PE experiments, a clean surface is obtained via \textit{in-situ} mechanical cleaving in the direction perpendicular to the atomic planes. The cleaving is done at 110~K, thus with the sample in the low-temperature monoclinic phase.

\subsection{Time- and angle-resolved, one- and two-photon photoelectron~spectroscopy}
We choose PE spectroscopy to investigate the quasiparticle nonequilibrium dynamics of \TNS\,with energy, momentum and time resolution, and perform one- and two-photon photoelectron measurements to obtain access to the occupied and unoccupied electronic band structure, respectively. Steady-state one-photon photoelectron measurements of the VBs are performed with femtosecond ultraviolet pulses at 6.22~eV, which is higher than the work function of \TNS, $\Phi$~=~5.40~eV~\cite{Thesis}. More challenging is the detection of the CBs that are unoccupied at equilibrium. To address them, two-photon photoelectron (2PPE) spectroscopy offers the most suitable approach. The 2PPE process relies on the absorption of two photons within the same femtosecond pulse having energy lower than the material work function: one photon excites an electron above the Fermi energy, \EF, into a CB, and the second one promotes it above the vacuum level, $E_{\text{vac}}$. In our study, we use photons at 5.28~eV.

\begin{figure}
\centering
\includegraphics[width=1\textwidth]{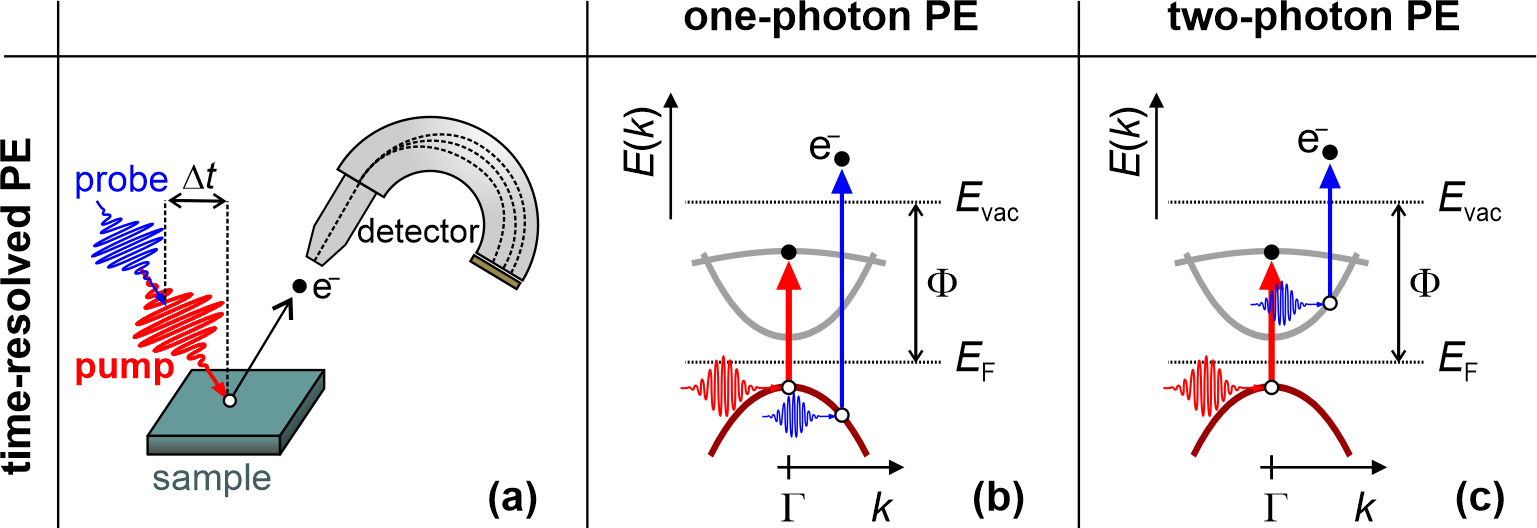}
		\caption[]{(a) Pump-probe experimental scheme for time- and angle-resolved, one- (b) and two-photon (c) photoelectron measurements of the occupied and unoccupied electronic band structure.}
	\label{Fig0b}
\end{figure}
To extend the measurements to the time domain, the pump-probe scheme depicted in \Fref{Fig0b}(a) is applied to both one- and two-photon photoelectron spectroscopy experiments. A ﬁrst intense pulse (’pump’) impinges on the sample and perturbs its equilibrium ground state by exciting valence electrons above the fundamental band gap into unoccupied CBs. After a time delay, $\Delta t$, a second pulse (’probe’) arrives on the sample and excites electrons above $E_{\text{vac}}$. In the tr-ARPES measurements (see \Fref{Fig0b}(b)), we use probe photons at 6.22~eV to enable direct PE from the top-most VBs dispersing around the $\Gamma$ point. In the tr-2PPE spectroscopy measurements (see \Fref{Fig0b}(c)), the probe photon energy is lowered to 5.28~eV to photoemit electrons only from the transiently occupied CBs dispersing around $\Gamma$. By variation of $\Delta t$, the transient occupation of the probed electronic bands is monitored in real time. 

Photoelectrons are collected by a hemispherical analyzer (SPECS Phoibos 100) as a function of their \SM{%\sout{kinetic energy, $E_{\text{kin}}$}
energy with respect to \EF}, and momentum, $k$. The 1.55~eV, 40 fs-long pump pulses are directly obtained from the fundamental output of a commercial regenerative amplifier system (\textit{Coherent RegA}). The probe pulses at 6.22~eV are generated from the amplifier fundamental output via a two-stage noncollinear sum-frequency-generation process. The probe pulses at 5.28~eV are generated in a two-stage fully-collinear optical parametric amplifier (\textit{Coherent}). In order to minimize the pump-induced heating of sample, we reduce the laser repetition rate to 40~kHz. In this way, we ensure control on the initial temperature of the sample, and thus, on its initial electronic and structural phase. Similar to what is discussed in our previous tr-ARPES work~\cite{Mor2017}, we check for the absence of surface band bending~\cite{Graf10}, and we employ sufficiently low pump fluences to prevent detrimental vacuum-space-charge effects on the spectra~\cite{Zhang12}. Energy resolution of 50~meV is estimated from the secondary-electron spectrum cut-off. Time resolution of 110~fs is obtained from the pump-probe pulses’ cross-correlation intensity. 
\section{Mapping of the occupied and unoccupied electronic states}
\label{sec:sec3}
\begin{figure}[t]
    \centering
    \includegraphics[width=0.6\textwidth]{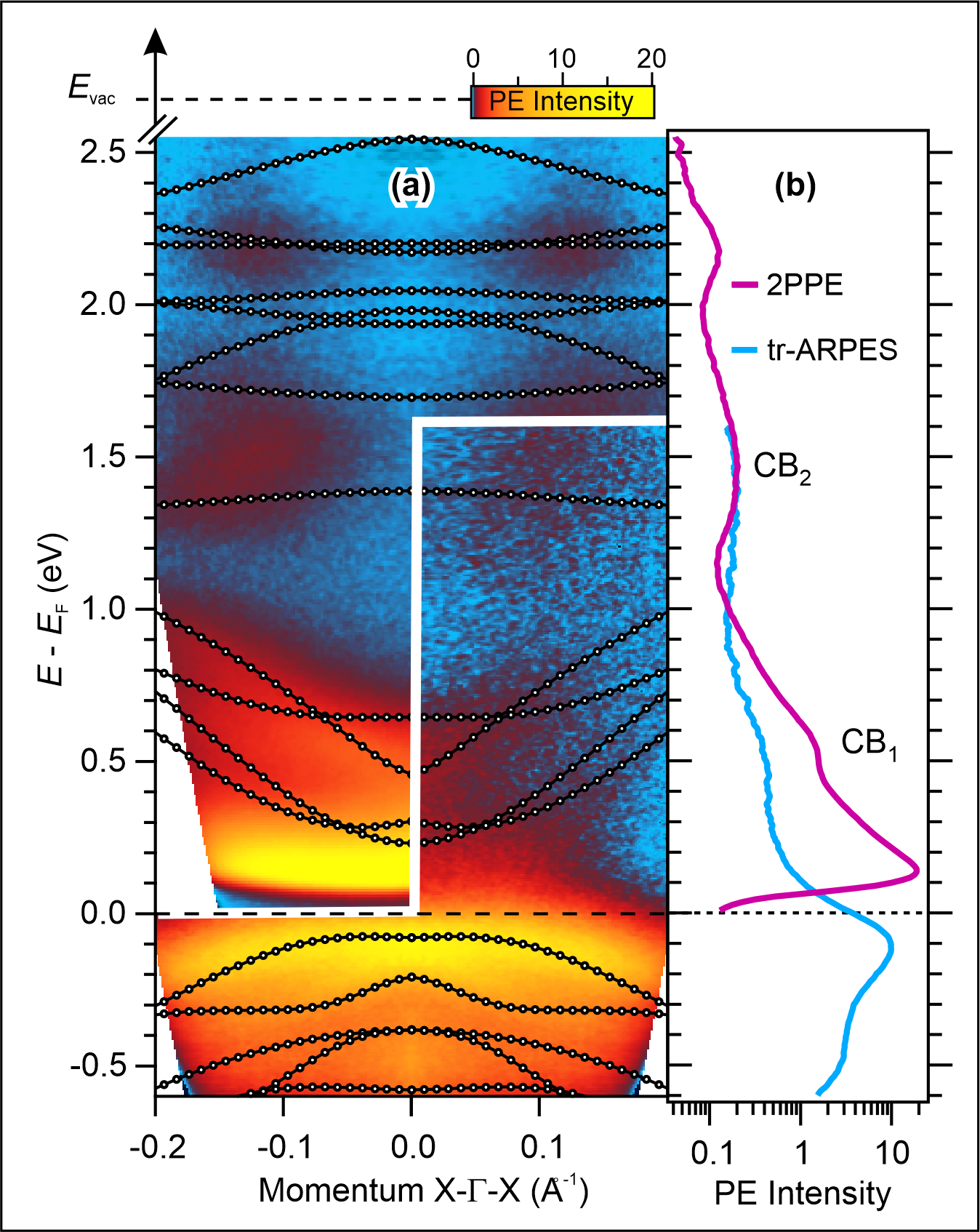}
    \caption{(a) Occupied and unoccupied electronic band structure of \TNS\,measured by tr-ARPES
    %(h$\nu_{pump}$~=~1.55~eV, h$\nu_{probe}$~=~6.22~eV)
    and steady-state 2PPE, %(2~$\times$~h$\nu$~=~5.28~eV)
    and DFT calculations from \cite{Monney2020}. To match the experimental band gap, the calculated CBs and VBs are shifted upwards by 200~meV and downwards by 50~meV, respectively. The spectrum below \EF\,and above 1.6~eV, and the calculated bands at all energies are symmetrized with respect to $\Gamma$. (b) Momentum-integrated EDCs of the data in (a). CB$_{\text{1}}$ and CB$_{\text{2}}$ label the low-energy, quasi-linearly dispersing CB manifold and the downward-dispersing CB at 1.5~eV, respectively.% whose ultrafast electron dynamics are discussed in Section~\ref{sec:sec4a} and \ref{sec:sec4b}, respectively.
    }
    \label{Fig1}
\end{figure}

\Fref{Fig1}(a) displays the occupied and unoccupied electronic band structure of \TNS\, measured at 110~K, i.e. below the critical temperature \TC\,of the semiconductor-to-EI phase transition. The dataset is obtained via PE along the $X-\Gamma$ direction of the Brillouin zone, which corresponds to the crystallographic $a$ axis parallel to the quasi-one-dimensional Ta and Ni atomic chains in real space (see \Fref{Fig0}(b)). The DFT calculations taken from \cite{Monney2020} are superimposed on the experimental data. 

The measured momentum-resolved spectrum of \Fref{Fig1}(a) is divided by a white contour into two regions corresponding to distinct PE schemes. The bottom and right part of the spectrum shows the tr-ARPES spectrum probed by 6.22~eV photons 80~fs after the arrival of the pump pulse. %at 1.55~eV.
The adopted two-color pump-probe PE scheme enables the measurement of the occupied electronic band structure and the low-energy portion of the unoccupied one dispersing around $\Gamma$ . The corresponding energy distribution curve (EDC) integrated over the full accessible momentum, i.e. from $k$ = 0 $\text{\AA}$ to $k$ = 0.18 $\text{\AA}$, is shown in \Fref{Fig1}(b) as light blue curve. Below \EF, we observe two VBs with maxima at $\Gamma$ and $E - E_{\text{F}} \sim-0.5$ and $-0.1$~eV, respectively. Both the momentum and energy positions of the two VBs agree with the shown calculations and with previous ARPES and DFT works \cite{Waki2009,Kane2013,Seki2014,Mor2017}. Particularly, the top of the upper VB exhibits the peculiar flat dispersion around $\Gamma$ which is typically assigned to the formation of an EI ground state~\cite{Waki2009,Waki2012,Kane2015, Seki2014}. Above \EF\,up to ca. 1~eV, a CB manifold (marked as CB$_{\text{1}}$ in \Fref{Fig1}(b)) shows quasi-linear upwards dispersion with minima at the $\Gamma$ point and at energies $E - E_{\text{F}} \sim$~0.5 and 0.3~eV. Approximately 1.5~eV above \EF, an isolated band (marked as CB$_{\text{2}}$ in \Fref{Fig1}(b)) appears.

The top-most and left part of the spectrum of \Fref{Fig1}(a) is obtained via steady-state 2PPE with 5.28~eV laser pulses, as described in Section~\ref{sec:exper}. The respective momentum-integrated EDC interval is shown in~\Fref{Fig1}(b) as purple curve. At intermediate state energies higher than 1.5~eV, the spectrum shows PE intensity corresponding to a set of %\SM{\sout{weakly-dispersing}} 
unoccupied electronic bands, as predicted by DFT.  Approximately 1.5~eV above \EF, CB$_{\text{2}}$ exhibits a %\SM{\sout{weak}} 
downwards dispersion \SM{around its maximum at $\Gamma$ with band mass of ca. 3 times the free-electron band mass, as obtained upon fitting of the relevant calculated band with nearly-free-electron parabolic dispersion}. Below 1.0~eV above \EF, CB$_{\text{1}}$ is resolved and confirms the quasi-linear dispersion towards the minimum at $\Gamma$ probed also with 6.22~eV photons. Eventually, below approximately 0.30~eV, the spectrum is dominated by the secondary-electron intensity background. %By comparison with the DFT calculations and with the background-free spectrum obtained by tr-ARPES at the corresponding energy and momenta, we note that this PE intensity overlays the spectral region where the lowest CB is predicted to have its minimum at $\Gamma$. 

Overall, we find a good agreement between the measured and calculated electronic bands. We retrieve all the main features of the occupied electronic band structure discussed in the literature~\cite{Waki2009,Waki2012,Seki2014,Kane2013}. Moreover, exploiting 2PPE spectroscopy allows us to experimentally address the momentum-resolved unoccupied electronic band structure at energies up to a few eV above \EF, a region that has not been explored so far.

\section{Ultrafast relaxation dynamics of quasi-free conduction electrons}
\label{sec:sec4}
We perform complementary tr-ARPES and tr-2PPE measurements below \TC\,in order to reconstruct the ultrafast dynamics of the quasi-free carriers and the excitons in the low-temperature phase of \TNS. In this section, we first investigate the electron dynamics resonantly induced at $\Gamma$, and focus on the impact of the band dispersion on the enabled electron scattering channels and the effect of the pump fluence on the transient occupation of the CB. Then, we examine the electron dynamics induced in the low-energy, quasi-linear dispersing CB$_{\text{1}}$, and explore their dependence on both the pump fluence and the excess energy of the electrons with respect to \EF. 

\subsection{Resonantly induced dynamics of the downwards-dispersing \texorpdfstring{CB\textsubscript{2}}{} at \texorpdfstring{$\Gamma$}{}}
\label{sec:sec4a}
\begin{figure}[t]
    \centering
    \includegraphics[width=0.5\textwidth]{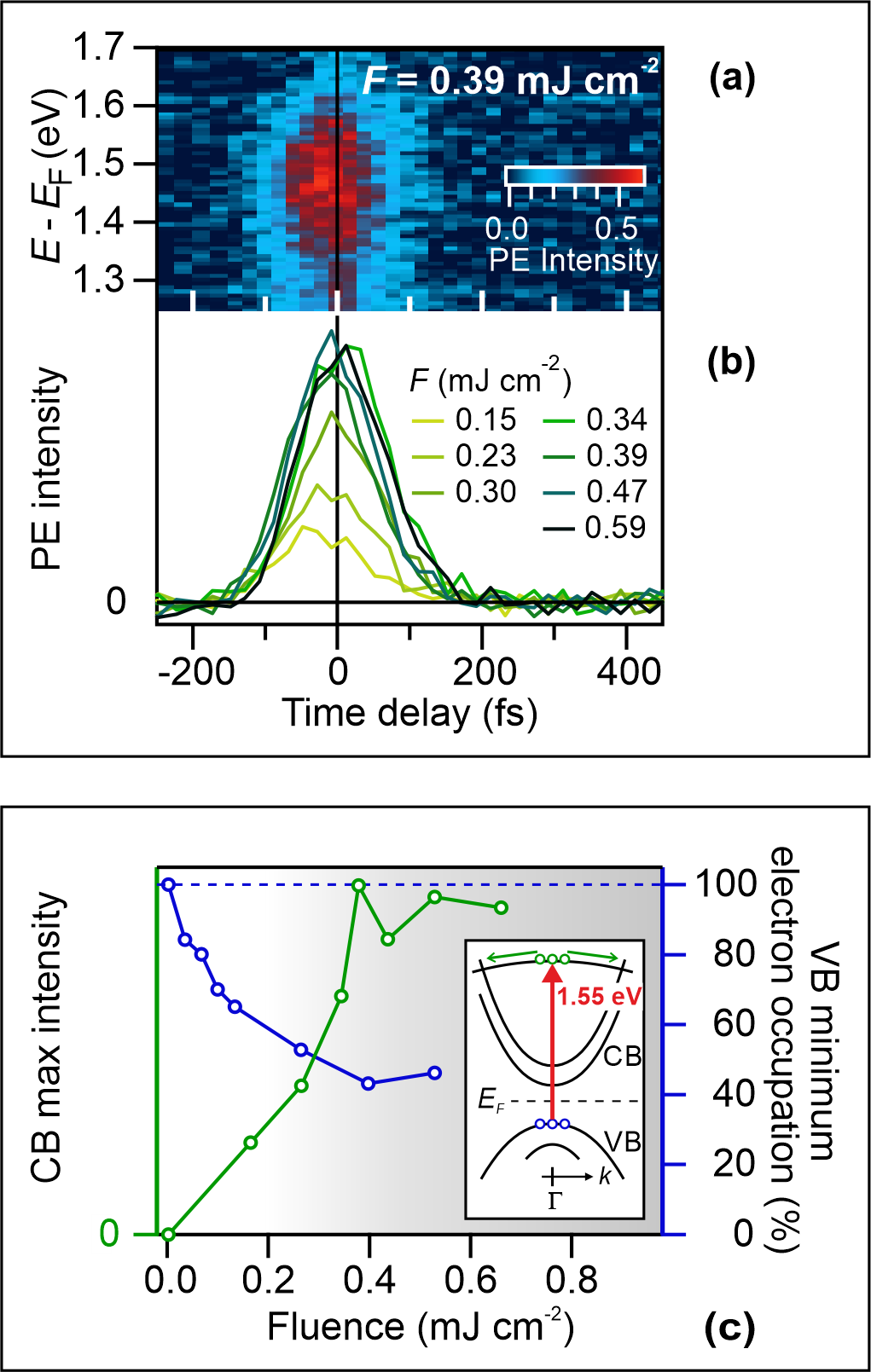}
    \caption{(a) Time- and energy-resolved PE intensity of CB$_{\text{2}}$ at 110~K. (b) Energy-integrated PE intensity for various pump fluences. (c) CB$_{\text{2}}$ occupation (left axis) and VB depopulation (right axis) as a function of pump fluence. Inset: schematic of the resonant excitation and subsequent scattering of electrons at $\Gamma$.}
    \label{Fig2}
\end{figure}

We use pump photons at 1.55~eV to resonantly induce electron dynamics  from the flat top of the upper VB to the downwards-dispersing CB$_{\text{2}}$ at $\Gamma$ (see red arrow in inset of \Fref{Fig2}(c)). %(cf. band structure and EDC shown in \Fref{Fig1}(a)-(b)). %These photons enable, at $\Gamma$, resonant transitions of electrons from the flat top of the upper VB to CB$_{\text{2}}$.
The resulting transient electron population, integrated over all the accessible \textit{k}-vectors around $\Gamma$ (i.e. over the momentum range shown in \Fref{Fig1}(a)), is reported in \Fref{Fig2}(a) in false-color scale as a function of intermediate state energy with respect to \EF\,(left axis) and pump-probe time delay (bottom axis). The data shows a spectral feature with maximal intensity at zero-time delay and ca. 1.5~eV, being approximately 200~meV broad. The intensity rapidly decreases and almost vanishes 100~fs after the pump arrival. The corresponding energy-integrated PE intensity is reported in \Fref{Fig2}(b) and exhibits, independent of the pump fluence, a cross-correlation limited temporal profile. This observation suggests that, on a time scale faster than 110 fs, electrons excited to CB$_{\text{2}}$ undergo intraband scattering towards momenta exceeding the range explored in our measurements. Now, this CB exhibits a downwards dispersion along the $\Gamma$-X direction (see \Fref{Fig1}(a)), as opposed to any other momentum directions along which the calculated electronic bands are found to be upwardly dispersing~\cite{Kane2013,Monney2020}. Given that, electrons resonantly excited to CB$_{\text{2}}$ at $\Gamma$ should preferably scatter along $\Gamma$-X in order to release their excess energy. 

To quantify the effect of the pump fluence on the transient occupation of CB$_{\text{2}}$, we analyze the time-integrated PE intensity of each dataset shown in \Fref{Fig2}(b). The integration results are reported on the left axis of \Fref{Fig2}(c) as a function of fluence (bottom axis). For comparison, the transient PE intensity minimum of the VB top at $\Gamma$ is shown on the right axis, as evaluated in our previous work~\cite{Mor2017}.% The latter intensity represents the transient electron occupation minimum of the initial state of the resonant transitions induced by the pump pulses into the downward-dispersing CB$_{\text{2}}$.

We observe that with increasing pump fluence, both the CB$_{\text{2}}$ occupation and the VB depopulation increase up to a saturation which is completed for fluence values above 0.4~mJ~cm$^{-2}$. As reported in Section~\ref{sec:intro}, the occurrence of a saturation threshold in the nonequilibrium response of \TNS\,has been reported by several experimental works~\cite{Mor2017,Mor2018,Werdehausen2018,Saha2021,Bretscher2021}. In particular, our previous tr-ARPES~\cite{Mor2017} and transient reflectivity~\cite{Mor2018} studies have shown that the effect is caused by absorption saturation of pump photons at $\Gamma$ due to half-depletion of the initial state (see the 50\% VB occupation in \Fref{Fig2}(c) above 0.2~mJ~cm$^{-2}$) in analogy with a photoexcited two-level system. The observation that the intensity of the transition's final state, i.e. the CB$_{\text{2}}$ population excited at $\Gamma$, also saturates at comparable fluence values is in good agreement and thus provides complementary demonstration of the occurrence of an optical absorption saturation of pump photons in \TNS. 

\subsection{Fluence- and energy-dependent dynamics in the quasi-linearly dispersing \texorpdfstring{CB\textsubscript{1}}{}}
\label{sec:sec4b}
\begin{figure}
    \centering
    \includegraphics[width=0.9\textwidth]{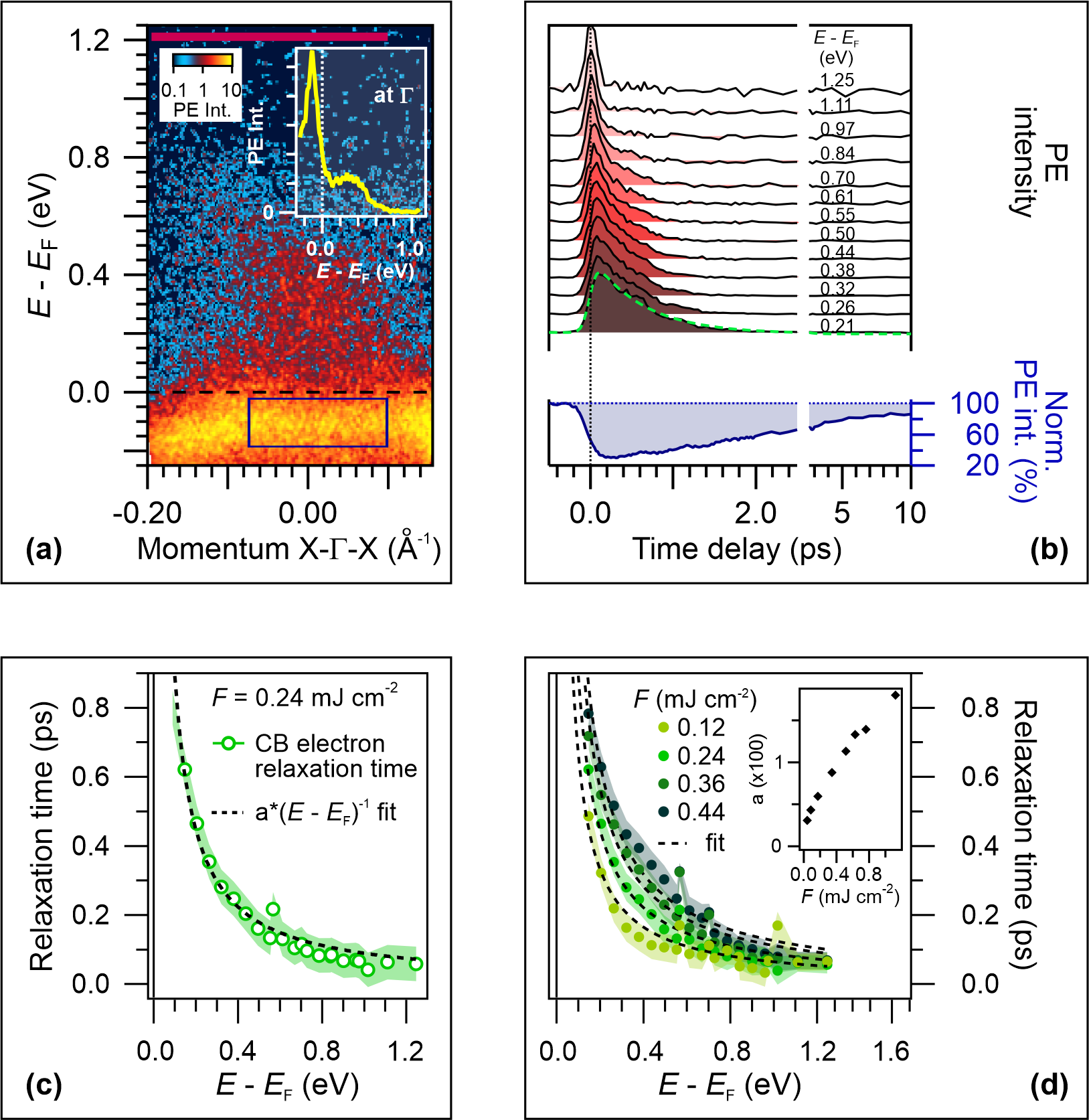}
    \caption{(a)~Momentum-resolved PE spectrum at 100~K, 50~fs and $0.44$~mJ~cm$^{-2}$ in logarithmic color scale. The red bar indicates the momentum interval of integration for the tr-PE intensities extracted at various intermediate energies in (b). Inset: EDC at $\Gamma$. (b)~Tr-PE intensity at indicated energies, offset for clarity. The dashed line is a monoexponential fit. Bottom panel: tr-PE intensity of the VB top normalized to the equilibrium intensity. (c)~CB$_{\text{1}}$ relaxation time as a function of intermediate energy and inverse linear fit. (d)~Energy-resolved CB$_{\text{1}}$ relaxation time for various fluences and best fits. In inset, the coefficient of proportionality as a function of fluence.}
    %\textcolor{red}{HERE A NATURAL QUESTION IS: HOW DO THE PROPORTIONALITY FACTORS SCALE WITH F? IS THERE A REASON NOT TO SHOW THESE?
    \label{Fig3}
\end{figure}
We now analyze the electron relaxation dynamics in the low-energy, quasi-linearly dispersing CB$_{\text{1}}$. This CB manifold is populated indirectly due to the relaxation of quasi-free carriers form higher-energy states, as well as, likely, directly at the pump-pulse arrival by vertical transitions at momenta exceeding the interval accessible by our probe photons. \Fref{Fig3}(a) shows the energy- and momentum-resolved snapshot of the CB$_{\text{1}}$ electron occupation 50~fs after photoexcitation with fluence of 0.44~mJ~cm$^{-2}$. Here, 6.22~eV probe photons are used in order to resolve the PE intensity at the CBM. First, we note that the CBM peaks at ca. 0.3~eV above \EF\,and is well separated from the VB top. This is emphasized by the EDC extracted at $\Gamma$ and shown in the inset (yellow curve). The absence of band crossing between CB$_{\text{1}}$ and the top-most VB at $\Gamma$ is consistent with our previous finding that a photoinduced closure of the electronic band gap is not achieved at this excitation density~\cite{Mor2017}, as opposed to other tr-ARPES works performed at much higher pump fluences~\cite{Okazaki2018,Saha2021}.

The transient CB$_{\text{1}}$ population is then investigated as a function of the excess energy with respect to \EF. The traces in \Fref{Fig3}(b) are the tr-PE intensities integrated over various energy intervals of few tens of meV, from $E$~-~\EF~=~1.25~eV down to 0.21~eV, and over the momentum interval indicated by the red bar in \Fref{Fig3}(a). To obtain this data set, photoelectrons at $E$~-~\EF~$>$~0.44~eV are probed by 5.28~eV photons, while at lower energies, 6.22~eV probe photons are used. At the bottom of panel (b), the transient electron occupation of the flat VB top also probed by 6.22~eV photons is shown for completeness (blue shade). This intensity is obtained upon integration in the energy-momentum region indicated by the blue box in \Fref{Fig3}(a). The tr-PE intensity at energies above \EF\,is found to decay on a sub-picosecond time scale. Also, the intensity decay slows down with decreasing excess energy. At the lowest excess energy of 0.21~eV, no transient intensity is recorded after ca. 2~ps. Conversely, the PE intensity of the VB at $\Gamma$ exhibits much slower recovery dynamics which is not fully completed even up to 10~ps.

To quantify the energy dependence of the relaxation dynamics of CB$_{\text{1}}$ electrons, we fit the transient PE intensity at each energy with a monoexponential decay function convoluted with a Gaussian crosscorrelation function, as exemplified by the green dashed line in panel (b). For an excitation with a pump fluence of 0.24~mJ~cm$^{-2}$, the relaxation time constant obtained from the fits is plotted in graph (c) as a function of the excess energy. We find that the CB$_{\text{1}}$ electron relaxation time scales linearly with the inverse excess energy as evidenced by the black dashed fit curve in graph (c) obtained with a proportionality constant $a$ as the only fitting coefficient. We then test the robustness of such inverse linear dependence on a set of time constants obtained upon fitting of the tr-PE intensity at various pump fluence values. The results are reported in graph (d). Clearly, the dependence is confirmed at all fluence values with the proportionality constant $a$ that increases with increasing fluence, i.e. the density of excited carriers, as reported in the inset.

The observed linear dependence on the inverse excess energy is uncommon for conduction electrons photoexcited in a bulk semiconductor. However, \SM{it is known that the electron-electron scattering depends on the density of states (DOS) of the material that varies, in turn, for  different dimensionality~\cite{Pet97} and for different dispersions of the involved bands~\cite{Kittel}. In the present case, we observe a quasi-linear dispersion of the CB above its flat region near $\Gamma$ (see \Fref{Fig1}) in agreement with band structure calculations of the layered compound \TNS~\cite{Monney2020,Kane2013}. Linear bands (e.g. acoustic phonon bands) exhibit a constant or linear DOS in one or two dimensions, respectively. %,containing chains of metal atoms suggests either a 2D or 1D character of the DOS, depending on the character of the band, which would lead to a linear or constant DOS as a function of energy. 
The observed linear energy dependence of the relaxation rate of electrons in the quasi-linearly dispersing CB is, thus, a clear indication of electron relaxation by intraband scattering within a two-dimensional band with linear dispersion~\cite{Zhang17}. This is in agreement with previous works of layered systems such as graphite~\cite{Gonzalez1996} and twisted bilayer graphene~\cite{Gonzalez2020} where the electron-electron scattering time scales linearly with the inverse excess electron energy.}
%Significantly, the electron-electron scattering time of layered systems such as} graphite~\cite{Gonzalez1996} and twisted bilayer graphene~\cite{Gonzalez2020} scales linearly with the inverse excess electron energy. 
Such dependence has been explained by a combination of the linear dispersion of the electronic bands in the vicinity of \EF\,and the reduced screening of the Coulomb interaction due to low concentration of mobile carriers that stem form the reduced dimensionality of these materials. Notably, both aspects are satisfied in \TNS\,which is characterized by a quasi-linear dispersion of CB$_{\text{1}}$ around its minimum (see \Fref{Fig1}(a)) and a \SM{%\sout{suppressed} 
weak} screening of the Coulomb interaction \SM{due to low concentration of mobile carriers}. We thus propose that the nearly linear band dispersion combined with the low concentration of mobile carriers of \TNS\,contribute to justify the peculiar energy dependence of the CB$_{\text{1}}$ electron relaxation time.

The fluence dependence of the relaxation time allows us to further infer on the effect of an increasing density of photoexcited quasi-free carriers on their scattering rate. Such rate is typically reduced by means of the transient enhancement of %\SM{\sout{Coulomb}} 
screening \SM{that reduces the Coulomb interaction between scattering partners }% by the \SM{surrounding} photoexcited quasi-free carriers
~\cite{Bauer2015}. Our experimental results reveal a fluence-dependent increase of the relaxation time (inset of \Fref{Fig3}(d)) consistent with the 
screening-induced reduction of scattering and the consequent delay of electron relaxation in the CB.
%Our experimental results reveal a slower relaxation dynamics of the CB electron population with increasing the pump fluence, i.e. with the density of photoexcited carriers. This fluence dependent behavior is thus consistent with the discussed screening-induced reduction of scattering processes and the consequent delay of electron relaxation in the CB.

To sum up, the relaxation of conduction  electrons around $\Gamma$ involves rich dynamics. Electrons resonantly excited to the downwards-dispersing CB$_{\text{2}}$ scatter through the band towards larger momenta. In real space, this corresponds to electron diffusion along the quasi-one dimensional Ta atomic chains. Within less than ca. 100~fs, the process leads to a vanishing PE intensity in the explored momentum range. The electron population excited to the quasi-linearly dispersing CB$_{\text{1}}$ exhibits sub-picosecond decay dynamics that slow down for both excess energy closer to \EF\,and stronger excitation density. We proposed that the nontrivial linear dependence of the relaxation time on the inverse excess energy are determined by the quasi-linear dispersion of CB$_{\text{1}}$ towards its minimum at $\Gamma$ and the suppressed screening of the Coulomb interaction. We hope that our experimental results will trigger theoretical studies aiming at a quantitative prediction of these energy-dependent electron relaxation dynamics in \TNS. Finally, we found that transient enhancement of the screening of the Coulomb interaction by photoexcited carriers slows down the relaxation of conduction electrons.

\section{Ultrafast probing of valence electrons and condensed excitons}
\label{sec:sec5}
\begin{figure}[t]
    \centering
    \includegraphics[width=1\textwidth]{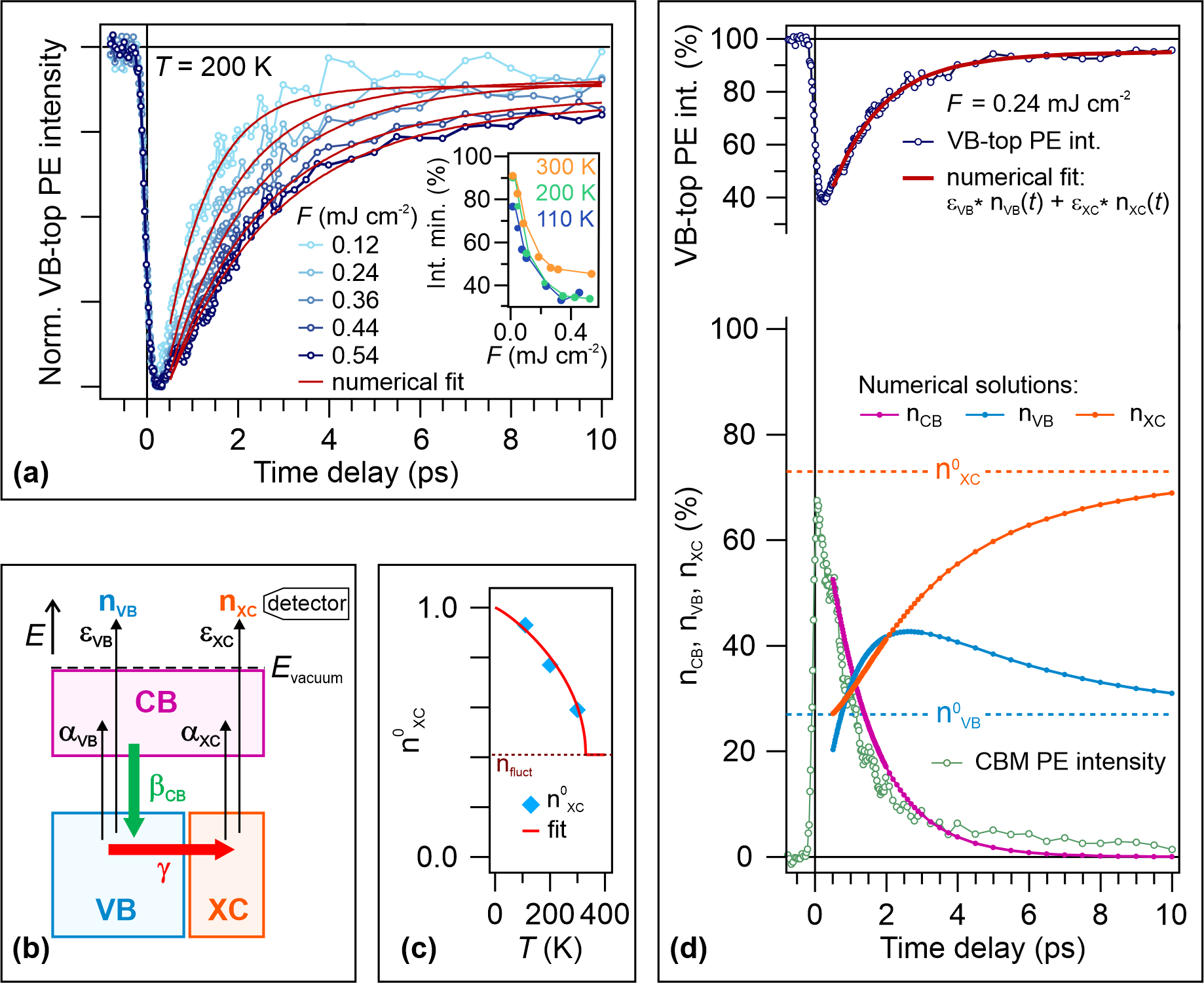}
    \caption{(a) Tr-PE intensity of the VB top (circles) at 200~K and numerical fits (red lines) for various pump fluences. Inset: intensity minimum as a function of fluence. (b) Schematic of the rate equation model. (c) Equilibrium XC occupation $n^0_{\text{XC}}$ as a function of temperature and fit as mean-field order parameter. (d) Exemplary numerical fit to the data and relevant solutions to the rate equations. See main text for details.}
    \label{Fig4}
\end{figure}
We now focus on the quasiparticle relaxation dynamics of the VB top. Particularly, we aim to shine light on a %the
question emerged from \Fref{Fig3}(b): how do we explain the slow ($>$10~ps) recovery dynamics of the VB intensity with the much more rapid ($<$2~ps) decay of the CB population? At a first glance, such discrepancy in the time scales seems to indicate that the total number of electrons is not conserved after photoexcitation within the probed region in real and momentum space.

One possible explanation is that some CB electrons diffuse away from the sample surface over a distance longer than the photoelectron escape depth before their decay across the band gap occurs. %~\cite{Retten97,Kirchmann10}.
As a result, the PE intensity decay of the CB is shortened and the PE intensity of the VB is lacking the contribution of the diffused electrons. In our experiment, the diffusion away from the surface corresponds to the transfer of electrons along the crystallographic $c$ axis perpendicular to the atomic planes of \TNS\,(see \Fref{Fig0}(b)) whose interlayer distance is 12.829~\r{A}~\cite{Sunshine1985}. At the used probe photon energy, the photoelectron escape depth is on the order of 5~nm~\cite{Huf96}. This length corresponds to 4-5 times the interlayer distance, and should be covered well within 2~ps by the CB$_{\text{2}}$ electrons in order to verify the proposed scenario. 
%, the CB electrons should travel away from the surface more than four times the interlayer distance within 2~ps. 
However, as discussed in Section~\ref{sec:sec3}, the calculated dispersion of CB$_{\text{2}}$ along the  $\Gamma$-Y momentum direction is very flat around $\Gamma$ and positively dispersing at larger momenta~\cite{Monney2020}. All this indicates a low electron mobility in the direction %of the electrons across the atomic planes away from
perpendicular to the surface as scattering through CB$_{\text{2}}$ is not energetically favorable along $\Gamma$-Y. In fact, this line of arguments would rather imply the opposite, i.e. the VB-top PE intensity decaying faster than the CB one, since the dispersion of the VB perpendicular to the surface is larger than that of the CB. Eventually, the absence of surface band banding (see Section~\ref{sec:exper}) rules out any driving force for transient surface photovoltage effects, such as an accelerated diffusion of electrons into the bulk. 
%we have ruled out the occurrence of transient surface photovoltage~\cite{Zhang12,Staehler17,Gierster21} in our measurements which allows us to exclude that the electron diffusion into the bulk may be accelerated by a transient electric field induced by the spatial separation of electrons and holes at the sample surface. 
Therefore, the scenario of a spatial diffusion of electrons outside the length scale of interest of our experiment appears unlikely or at least not sufficient to justify the significant mismatch of the CB and VB intensities. Because the detected CB manifold constitutes the lowest-energy unoccupied electronic bands of \TNS, we further exclude that the CB electrons may vanish from the momentum interval accessible by our probe photons due to scattering into bands at larger wave vectors. In fact, the CB population is expected to collect at $\Gamma$ upon release of the excess energy before recombining with the VB holes.%\textcolor{red}{IN THIS PARAGRAPH, PLEASE ALSO PUT A BRIEF COMMENT WHY THE ELECTRONS CANNOT JUST VANISH IN MOMENTUM SPACE ALSO, BECAUSE THERE ARE NO BANDS AT LOWER ENERGY (ARE THERE?)}

In the following, we show that the distinct temporal evolution of the CB and VB intensities can be reconciled by considering the origin of the VB-top PE intensity in the presence of an EI phase. Specifically, while the PE intensity of the CB simply stems from a photoexcited population of quasi-free electrons, the signal of the VB top is more composite in nature. To set the background for the discussion, we remind the reader that, at equilibrium below \TC, the VB top of \TNS\,is expected to host a condensate of ground state excitons which co-exist with the valence electrons. Consequently, two concurrent PE processes can take place at the VB top: the emission of valence electrons, and the breaking of condensed excitons by the PE of electrons. To be able to distinguish the two contributions to the photoelectron population, we extend the investigation of the VB-top PE intensity to the time domain, where the evolution of the two quasiparticle sets (VB and XC) may show different dynamics.

\subsection{Ground state exciton re-formation: bottleneck for the VB population recovery}
\label{sec:sec5a}
The tr-PE intensity of the VB top at 200~K is shown in \Fref{Fig4}(a) for various pump fluences (light-blue to blue circles). Each dataset is normalized to the intensity at negative delays in order to reproduce the transient band occupation, and is scaled to match the minima in order to emphasize the fluence dependence of the intensity recovery dynamics. We observe the following: at all pump fluences, the VB intensity exhibits an abrupt suppression followed by a multistep recovery that temporally exceeds the measured time delay of 10~ps; the intensity recovery dynamics increasingly slow down for higher fluence values. Moreover, as shown in the inset of \Fref{Fig4}(a), the abrupt intensity suppression scales linearly with increasing fluence up to the expected optical absorption saturation threshold~\cite{Mor2017,Mor2018}. Also, this behavior is robust against temperature changes below \TC.

Aiming for a full reconstruction of the ultrafast relaxation dynamics of quasiparticles in \TNS, we develop a rate equation model that is able to rationalize the multistep recovery dynamics of the VB-top PE intensity for all explored fluences and temperatures below \TC. The model assumes that electrons, excitons and phonons have thermalized to a common transient temperature that does not significantly change in the accounted time window. %As such, the fit parameters, $\beta_{\text{CB}}$ and $\gamma$, which could in principle depend on the transient electronic temperature, are considered to be time-independent. 
From the photoinduced dynamics of the EI band gap reported in our previous tr-ARPES study, we estimate that thermalization occurs after approximately 0.5~ps~\cite{Mor2017}. This value corresponds to the time delay at which the nonequilbrium EI band gap enhancement is completed at high fluences, and the thermal-induced band gap shrinking sets in. Accordingly, the model is fitted to the data points at time delays later than 0.5~ps. A scheme of the model is represented in \Fref{Fig4}(b). We consider three sets of quasiparticle populations, namely the quasi-free CB electrons, $n_{\text{CB}}$, the VB electron population, $n_{\text{VB}}$, and the electrons bound to holes in the XC, $n_{\text{XC}}$. We assign a distinct probability to the VB electrons, $\alpha_{\text{VB}}$, and the XC electrons, $\alpha_{\text{XC}}$, to absorb a pump photon, and we assume that both processes result in the excitation of a quasi-free electron to the CB, thus indistinctly on the initial state of the photoinduced electronic transition. We then introduce distinct PE probabilities, $\epsilon_{\text{VB}}$ and $\epsilon_{\text{XC}}$, for the respective sets, to photoemit an electron upon absorption of a probe photon. We note that in the case of a photoelectron emitted from the XC, this additionally implies the breaking of a bound electron-hole pair. 

To describe the relaxation dynamics, the model accounts for the recombination of the quasi-free CB electrons with the holes of the VB with a rate $\beta_{\text{CB}}$, as represented by the green arrow. This decay process is responsible for the emptying of the CB and the re-filling of the VB. The equilibrium XC occupation is re-established with a rate $\gamma$ (red arrow) through the re-formation of ground state excitons by VB electrons. We highlight that this model aims to describe the dynamics of a closed system with no electron loss at any time. %Also, we assume that the model holds at time delays where electrons, excitons and the lattice have thermalized to a common transient temperature that does not significantly change in the accounted time window. As such, the fit parameters, $\beta_{\text{CB}}$ and $\gamma$, which could in principle depend on the transient electronic temperature, are considered to be time-independent. From our previous time-resolved ARPES study, we estimate a thermalization time of approximately 0.5~ps~\cite{Mor2017}. Accordingly, the model will be employed to fit data points at time delays later than 0.5~ps.

The depicted model can be expressed by the following coupled differential equations, which account for the transient occupation of each quasiparticle set:
\begin{align}
    &\dot{n}_{\text{CB}} (t) = -\beta_{\text{CB}}~ n_{\text{CB}} (t)\\
%\end{equation}
%\begin{equation}
    &\dot{n}_{\text{VB}} (t) = ~~\beta_{\text{CB}}~ n_{\text{CB}} (t) - \gamma~ n_{\text{VB}}(t)~(n_{\text{XC}}^0 - n_{\text{XC}}(t))\\
%\end{equation}
%\begin{equation}
    &\dot{n}_{\text{XC}} (t) = ~~~~~~~~~~~~~~~~~~~~~\gamma~ n_{\text{VB}}(t)~(n_{\text{XC}}^0 - n_{\text{XC}}(t))
\end{align}
where $\beta_{\text{CB}}$ is the decay rate of quasi-free CB electrons across the band gap, $\gamma$ the ground state excitons re-formation rate, and $n_{\text{XC}}^0$ the XC occupation at equilibrium. To satisfy the conservation of the total quasiparticle number, we set $n_{\text{CB}}^0 = 0$ and $n_{\text{XC}}^0$~+~$n_{\text{VB}}^0$~=~1 at equilibrium, i.e. for $t\to\infty$, and the initial constraint on the CB population as given by the relationship: $n_{\text{CB}}(t = 0^{+}) = \alpha_{\text{VB}}n_{\text{VB}}^0 + \alpha_{\text{XC}}n_{\text{XC}}^0$, where $t = 0^{+}$ corresponds to the earliest time delay accounted by in the fitting of the data, i.e. 0.5~ps. The value of $n_{\text{CB}}(t = 0^{+})$ changes with varying the pump fluence as $\alpha_{VB}$ and $\alpha_{\text{XC}}$ depend on the excitation density. Accordingly, the initial VB and XC populations are $n_{\text{VB}}(t = 0^{+}) = n_{\text{VB}}^0(1-\alpha_{\text{VB}})$ and $n_{\text{XC}}(t = 0^{+}) = n_{\text{XC}}^0(1-\alpha_{\text{XC}})$, respectively.

Based on the described model, we express the transient PE intensity of the VB top, which encloses the entwined signatures stemming from the valence electrons and the photo-ionized ground state excitons, as a linear combination of two contributions, i.e. $I_{\text{VB}}(t) = \epsilon_{\text{VB}}n_{\text{VB}}(t) + \epsilon_{\text{XC}}n_{XC}(t)$. This expression is used to numerically fit globally the whole dataset of tr-PE intensity of the VB top recorded at three different temperatures below \TC\,and for various pump fluence values.% Through this approach, the contribution of the quasi-free VB electrons and the condensed excitons to the PE intensity can be eventually disentangled.  

In \Fref{Fig4}(a), we report the best fits to the dataset at $T$~=~200~K which exemplarily shows the very good agreement between the experiment and the numerical model. Specifically, the fits nicely reproduce the early recovery dynamics as well as the intensity depletion persisting on the long time scale. We stress that the fit function succeeds in fitting a large set of data with four free parameters, i.e. the two decay rates, $\beta_{\text{CB}}$ and $\gamma$, and the two excitation probabilities, $\alpha_{\text{VB}}$ and $\alpha_{\text{XC}}$. Additionally, three parameters have been globally fixed during the fitting, i.e. the equilibrium occupation of the XC, $n_{\text{XC}}^0$, that we let vary with temperature (while being hold with varying the pump fluence), and the two PE probability, $\epsilon_{\text{VB}}$ and $\epsilon_{\text{XC}}$, that we assumed constant with both temperature and fluence. 

We now look in detail into the fit results, and particularly, the numerical solutions of the model. The top graph of \Fref{Fig4}(d) shows an exemplary numerical fit curve (red curve) to the transient VB intensity~(blue circles) recorded at 200~K, after excitation with 0.24~mJ~cm$^{-2}$. %Overlaid on the same data, we also show an attempted fit by a monoexponential function (light blue) which clearly fails in following the late recovery dynamics. This comparison is to emphasize once more that a more composite model is indeed required to fit the experimental data.
The bottom part of \Fref{Fig4}(d) reports the numerical solutions of the three rate equations contributing to the fit of the transient VB-top intensity shown on the top part of the graph. We find that:
\begin{itemize}
    \item{The top-VB electron population $n_{\text{VB}}(t)$ (blue) starts from a modestly lower initial value than the equilibrium one $n^0_{\text{VB}}$ (blue dashed line), and exhibits a non-monotonic temporal evolution: at first, it rapidly increases up to largely overshooting $n^0_{\text{VB}}$ and reaching a maximum intensity at ca. 2~ps. Subsequently, $n_{\text{VB}}(t)$ slowly decreases, yet not fully recovering the $n^0_{\text{VB}}$ value after 10~ps.} 
    \item{The XC electron population $n_{\text{XC}}(t)$ (orange) is massively reduced with respect to the equilibrium occupation of the XC $n^0_{\text{XC}}$ (orange dashed line), and then continues to increase towards $n^0_{\text{XC}}$ across the explored time-delay interval.}
    \item{The excited electrons in the CB $n_{\text{CB}}(t)$ (purple) shows a rapid decay towards almost-zero occupation of the CB within approximately 2~ps, thus much more rapidly than the recovery of both the VB and XC populations.}
\end{itemize} 

In order to test the validity of the model, we compare the evolution of the numerical solution of the CB electron population $n_{\text{CB}}(t)$ with the %tr-2PPE 
data acquired at the CBM (green circles). Notably, the $n_{\text{CB}}(t)$ dynamics nearly coincide with the tr-PE intensity of the CBM. This remarkable agreement proves the robustness of the developed model.

In the following, we focus on the details of the temperature and fluence dependencies of the global fit parameters. \Fref{Fig4}(c) shows that the $n_{\text{XC}}^0$ decreases for temperatures approaching \TC\,in a manner that follows the mean-field evolution of an order parameter, i.e.
%\begin{center}
\begin{equation}
    n_{\text{XC}}^0(T) = (n_{0} - n_{\text{fluct}})*\sqrt{(1 - T/T_C)} + n_{\text{fluct}} 
\end{equation}
%\end{center}
with $n_0 = 1$, and \TC\, fixed at 328~K. The parameter $n_{\text{fluct}}$ describes a constant background of fluctuating excitons beyond a typical mean-field behavior, in agreement with previous work by Seki \textit{et al.}~\cite{Seki2014}. The value of 0.38 is quite high, but similar to what seen for TiSe$_2$~\cite{Monn2010}. $n_{\text{fluct}}$ is an \textit{ad-hoc} variable used to capture the fluctuation regime above \TC\,beyond a mean-field temperature model. %Such fluctuations have been shown to also exist above \TC\,in \TNS~\cite{Seki2014}.

Concerning the PE probabilities, we globally obtain $\epsilon_{\text{VB}} = 0.84$ and $\epsilon_{\text{XC}} = 1$, respectively. These values interestingly reveal, in the low-temperature phase on \TNS, a lower probability to emit an electron from the VB top compared to ionizing a ground state exciton, which is likely a consequence of different transition dipole moments to the final state in vacuum.

The obtained numerical solutions and fit parameters allow for several conclusions. First, the model ensures that all the photoexcited electrons are distributed across the CB, the VB and the XC without losses, and thus, the total number of quasiparticles is conserved in the system. Particularly, at time delays where $n_{\text{CB}}(t)$ is almost zero, the VB is found to transiently host an excess of electrons compared to its equilibrium occupation, while the XC is still massively depopulated. On later time scales of several picoseconds, the equilibrium occupation of the VB and the XC is re-established through the re-formation of ground state excitons from the excess VB electrons. As a second result, we find that the apparent occupation imbalance observed around the VB and the CB energies
%\textcolor{red}{around the VB and the CB energies (I WANTED TO FORMULATE MORE CAREFULLY, BECAUSE IN THIS SENTENCE, VB INCLUDES THE XC WHILE IN THE PREVIOUS ONE YOU DISENTANGLE XC AND VB}
at time delays larger than ca.~2~ps results simply from the lower probability to photoemit, and thus, to detect electrons in the VB that have not yet bound with holes, %, which is actually over-populated, 
compared to those that have already re-formed ground state excitons in the XC. Finally, we unveil the re-formation of ground state excitons as the bottleneck process for the recovery of the equilibrium occupation of each quasiparticle set.

\subsection{Estimate of the ground state exciton re-formation energy}
\label{sec:sec5B}
\begin{figure}
    \centering
    \includegraphics{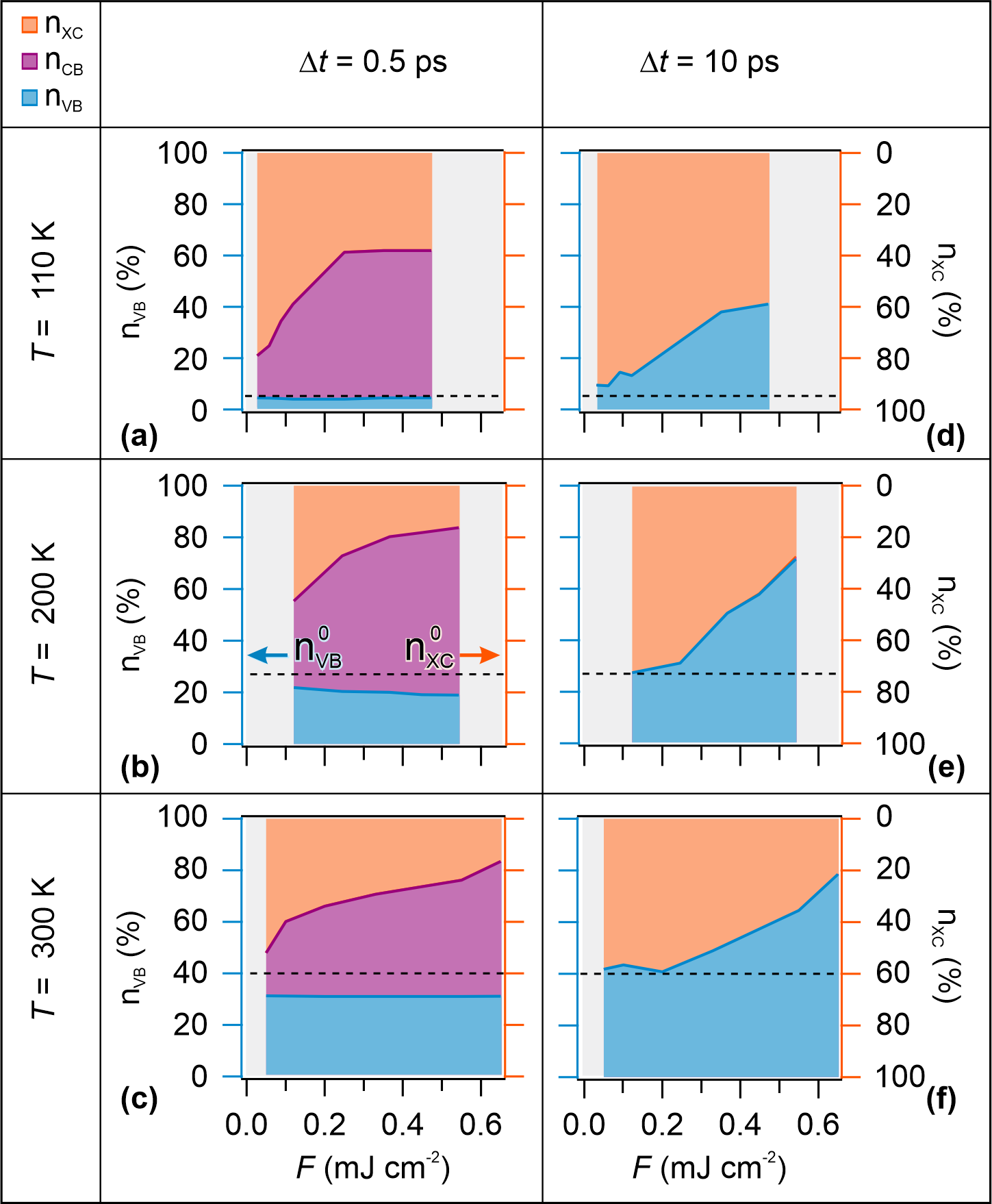}
    \caption{(a)-(f) The solutions $n_{\text{XC}}$ (orange), $n_{\text{CB}}$ (purple), and $n_{\text{VB}}$ (blue) as a function of pump fluence, as obtained for the datasets recorded at $T$~=~110~K (top), 200~K (middle), and 300~K(bottom). The solutions are evaluated at 0.5~ps (left) and 10~ps (right). The dashed black line indicates the fit parameter value of the equilibrium occupation of the VB $n_{\text{VB}}^{0}$ (left axes) and the condensate $n_{\text{XC}}^{0}$ (right axes) for each sample temperature.}
    \label{Fig6}
\end{figure}
In the previous subsection, we have shown that the PE intensity of the VB top is a good marker of the occupation of the XC below \TC\,in \TNS. Particularly, its temporal evolution allowed us to reconstruct the multistep relaxation pathway of the photoexcited electrons. The latter includes the re-formation of ground state excitons which is responsible for the retarded recovery of the equilibrium state. In the following, we aim at detailing how the ground state exciton re-formation evolves. Thereby, we will extract the small energy barrier which determines the rate $\gamma$ of this relaxation process.

\Fref{Fig6}(a)-(f) summarize the solutions $n_{\text{XC}}$ (orange), $n_{\text{CB}}$ (purple), and $n_{\text{VB}}$ (blue) as a function of pump fluence, as obtained for the datasets recorded at $T$~=~110~K (top), 200~K (middle), and 300~K(bottom). The solutions are evaluated at $\Delta t$~=~0.5~ps (left) and 10~ps (right), i.e. at the starting and ending points of the fits, respectively.

In each graph, the color-shaded distribution allows to readily grasp the transient occupation of each quasiparticle set compared to the whole quasiparticle population. We observe that, at $\Delta t$~=~0.5~ps, the XC occupation $n_{\text{XC}}$ (orange, right axes) is suppressed by several tens of percentage with respect to the equilibrium value $n_{\text{XC}}^0$ (dashed lines), and the suppression increases with increasing fluence up to a saturation. The VB population $n_{\text{VB}}$ (blue, left axes) remains quite close to its equilibrium value, and does not significantly change with fluence. %Actually, at the lowest initial sample temperature of $T$~=~110~K, $n_{\text{VB}}$ overshoots the equilibrium value, a situation which is occurring also at the higher temperatures at slightly later times (see e.g. the numerical solution $n_{\text{VB}}$($t$) at $T$~=200~K in \Fref{Fig4}(d)). 
Eventually, the CB occupation $n_{\text{CB}}$ manifests as the complementary part to $(n_{\text{VB}}$~+~$n_{\text{XC}})$ at each fluence value. We note that the higher the fluence, the larger $n_{\text{CB}}$, as expected. 

At the late time delay of $\Delta t$~=~10~ps, the population distribution into the three quasiparticle sets changes drastically: $n_{\text{XC}}$ is still lower than the equilibrium value particularly at higher fluence values; $n_{\text{VB}}$ has almost reached the equilibrium value at low fluences, while it strongly deviates from $n_{\text{VB}}^{0}$ for high fluences; no residual $n_{\text{CB}}$ intensity is observed. This redistribution of electrons into the three quasiparticle sets holds similarly at all temperatures investigated.

We now discuss the above model results, and relate them to the fit parameters. The fluence dependence of $n_{\text{XC}}$ at $\Delta t$~=~0.5~ps reflects the occurrence of pump-photon absorption saturation at the VB top for strong excitations~\cite{Mor2017}, and the consequent saturation of the VB-top PE intensity minimum shown in the inset of \Fref{Fig4}(a)). The increase of $n_{\text{CB}}$ with increasing fluence at 0.5~ps is consistent with the lower relaxation rate of CB electrons for stronger excitations measured by tr-PR spectroscopy of the excited states (see Subsection~\ref{sec:sec4a}). Accordingly, the VB re-filling rate, $\beta_{CB}$, decreases at higher pump fluences. At $\Delta t$~=~10~ps, the lack of $n_{CB}$ is congruous with the vanishing PE intensity of the CBM at this late time delay. On the other hand, the overshooting $n_{VB}$ compared to the equilibrium value indicates that the VB top still hosts an excess of electrons that have not bound with holes in order to recover the equilibrium XC occupation $n_{XC}^0$. The increasing discrepancy between $n_{VB}$ at $\Delta t$~=~10~ps and $n^0_{VB}$ with increasing fluence suggests, once again, that the re-formation rate of ground state excitons, $\gamma$, lowers for strong excitation. Accordingly,  the thermalization of the low-temperature EI phase is delayed.

\begin{figure}
    \centering
    \includegraphics[width=0.9\textwidth]{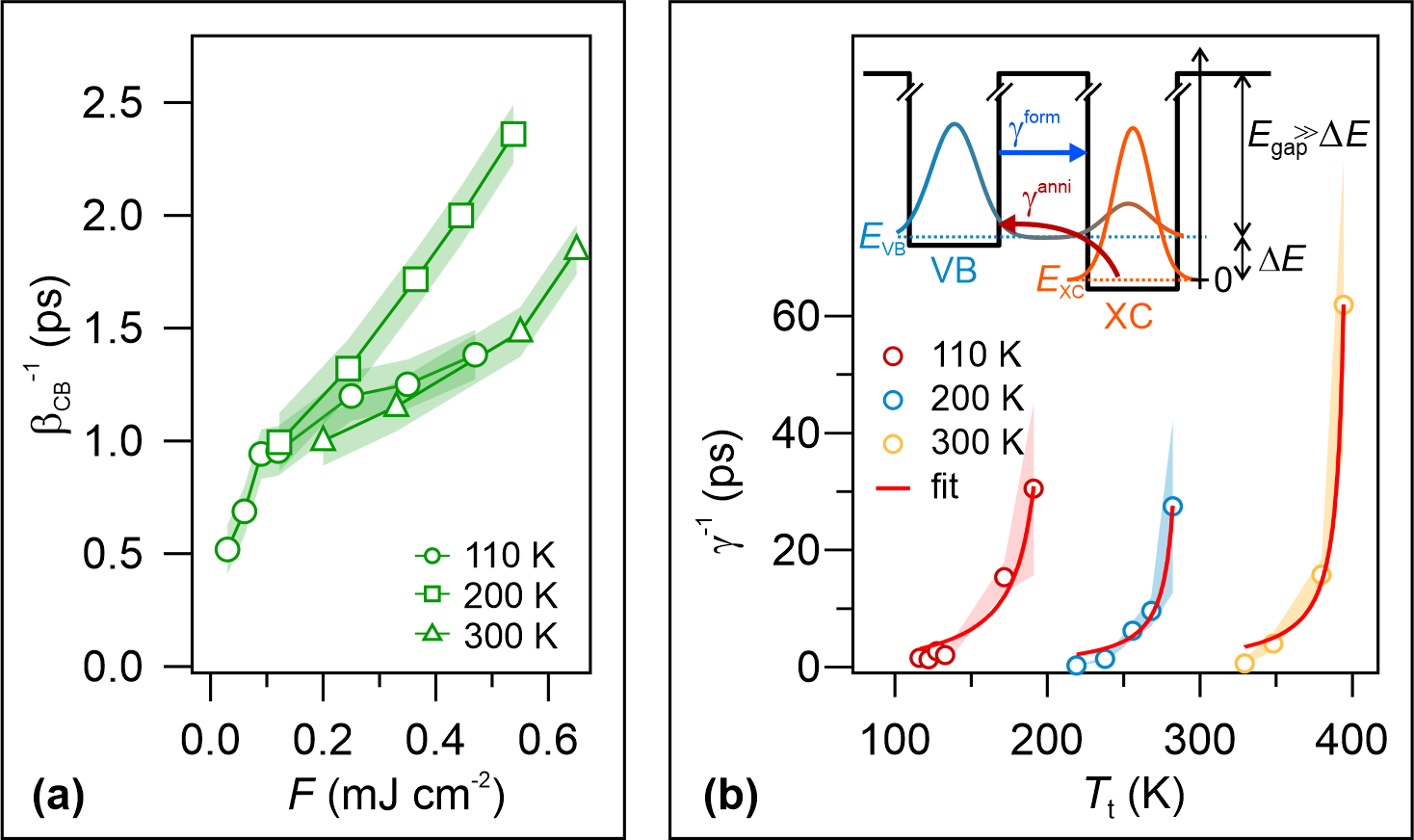}
    \caption{(a) The CB depopulation time $\beta^{-1}_{\text{CB}}$ as a function of pump fluence for different sample temperature. (b) The XC re-formation time $\gamma^{-1}$ as a function of transient temperature, $T_{\text{t}}=T_{\text{ini}}+\Delta T$, with $T_{\text{ini}}$ the initial sample temperature and $\Delta T$ the pump-induced increase. Inset: schematic of the two-quantum-well model (see main text for details).}
    \label{Fig7}
\end{figure}
As evidenced by the previous graphs, both the VB re-filling and the ground state exciton re-formation slow down with increasing pump fluence. To quantify these observations, we report the inverse fit parameters $\beta^{-1}_{\text{CB}}$ and $\gamma^{-1}$ providing the timescale of the two processes in \Fref{Fig7}(a) and (b), respectively. We find that the CB depopulation time, $\beta^{-1}_{\text{CB}}$, is on the order of 1~ps and increases with increasing pump fluence. We recall that the CB relaxation time independently obtained upon fitting the tr-PE data at the lowest accessible energy above \EF\,spans from 0.5~to~0.8~ps in the explored fluence range  (see \Fref{Fig3}(d)). Hence, $\beta^{-1}_{\text{CB}}$ differs only by a factor of ca. 2 despite the CB intensity not being fitted by the numerical model. We thus consider this result a pretty good agreement supportive of the validity of the numerical model. \SM{%\sout{From the fluence-dependence of $\beta^{-1}_{\text{CB}}$, it seems likely that the free-carrier-enhanced screening of the Coulomb interaction is responsible for retarding the electron-hole recombination across the band gap, in addition to the screening-induced slowing down of the CB electron relaxation though intraband scattering processes discussed in Subsection~\ref{sec:sec4b}} 
The increase of $\beta^{-1}_{\text{CB}}$ with increasing pump fluence likely reflects the transient enhancement of the screening of the Coulomb attraction between CB electrons and VB holes resulting in slower recombination~\cite{Cui2014,Silkin2015}. Furthermore, we observe an increase of $\gamma^{-1}$, the ground-state reformation time, with increasing quasiparticle density. This can also be rationalized by reduced Coulomb interaction, which is the driving force behind the reformation of ground-state excitons in the condensate. The quasiparticle screening increasingly hampers the restoration of the equilibrium XC occupancy with rising quasiparticle density.}
%with increasing pump fluence likely indicates the occurrence of a transient enhancement of screening of the Coulomb interaction between electrons and holes by means of a density of photoexcited quasi-free carriers: The larger the number of screening quasiparticles, the weaker the electron-hole Coulomb attraction~\cite{Cui2014,Silkin2015}. As a result exciton recombination across the band gap is hindered and its rate reduces, in analogy with} the screening-induced slowing down of the CB electron relaxation though intraband scattering processes discussed in Subsection~\ref{sec:sec4b}.

The ground state exciton re-formation occurs on a time scale $\gamma^{-1}$, which is on the order of up to tens of ps. Note that, at these late time delays, most of the absorbed energy has been converted to heat. It is, thus, interesting to evaluate how the exciton re-formation time may be affected by the photoinduced transient heating of the system. To do that, we calculate the pump-induced temperature increase, $\Delta T$, with respect to the initial temperature, $T_{\text{ini}}$, for each fluence value (see \ref{AppA}), and plot $\gamma^{-1}$ as a function of the transient temperature, $T_{\text{t}}=T_{\text{ini}}+\Delta T$. As shown in (b), at all initial temperatures, the ground state exciton re-formation time, $\gamma^{-1}$, increases upon transient heating of the system, and follows a strongly non-linear trend. Particularly, it tends to diverge when $T_{\text{t}}$ exceeds the initial temperature by 60-80~K. This is a remarkable behavior evidencing the occurrence of a process that counteracts the recovery of the equilibrium XC occupation at high transient temperatures. We propose that annihilation of ground state excitons may occur, and become more and more significant for higher transient temperatures. While the condensate is not destroyed even when $T_{\text{t}} >$~\TC, the annihilation of ground-state excitons into unbound electrons and holes slows down the recovery of the equilibrium occupations $n_{XC}^0$ and $n_{VB}^0$. In this scenario, $\gamma$ is interpreted as the effective rate resulting from two opposite processes , i.e.
\begin{equation} \label{eq:5}
\gamma = \gamma^{\text{form}} - \gamma^{\text{anni}}.
\end{equation}
For $\gamma > 0$, the re-formation of ground state excitons dominates over their annihilation, ensuring the recovery of the XC population $n^{0}_{XC}$, and thus, of the equilibrium PE intensity of the VB top of \TNS. With increasing transient temperature due to stronger photoexcitation, the exciton annihilation becomes more prominent and causes a reduction of $\gamma$. Eventually, we note that independently of the initial temperature of the system, $\gamma^{-1}$ exhibits a diverging behavior at a certain transient temperature. %$T^*$. 
We interpret this point as the condition when the electron backward process exactly compensates the forward one. This transient balance between creation and annihilation of ground state excitons is obtained when $\gamma^{\text{form}}$ equals $\gamma^{\text{anni}}$, and the effective rate $\gamma$ vanishes, accordingly. We point out that this condition does not imply the melting of the condensate, but rather describe a transient state towards equilibrium of the EI phase. 

%\textcolor{red}{I FELT THAT THE FOLLOWING BECAME TOO LONG AND TOO BIG. I SUGGEST TO CHANGE IT AS FOLLOWS:}
%To test this hypothesis, we develop a phenomenological model to fit $\gamma^{-1}$ as a function of $T_{\text{t}}$. A schematic of the model is depicted in the inset of \Fref{Fig7}(b). The two quantum wells correspond to the energy levels of the quasiparticle sets of electrons in the VB and the XC, respectively. Accordingly, the bottom of each well defines the energy of an electron at the VB top, $E_{\text{VB}}$, or bound to a hole in the XC, $E_{\text{XC}}$. The difference $\Delta E = E_{\text{VB}}-E_{\text{XC}}$ gives the energy gain in forming a ground state exciton in the EI phase. As such, we expect $\Delta E\ll1$ at finite temperatures $T <$ \TC, to account for the predicted spectral vicinity of the VB top and the excitonic ground state (see Section~\ref{sec:intro}). The blue arrow labeled as $\gamma^{\text{forward}}$ represents the re-formation of ground state excitons in the XC helping the relaxation of the system towards equilibrium, while the red arrow labeled as $\gamma^{\text{forward}}$ the exciton annihilation counteracting the EI phase re-establishment.

In a simplified picture, the transition of an electron from the VB to the XC can be viewed as a quantum tunnelling process, not 
%as tunnelling 
in real space, but 
%tunneling 
from one quantum state of the system to the other, as illustrated by the double-well schematic in \Fref{Fig7}(b). The height of the VB well, i.e. the fundamental electronic band gap $E_{\mathrm{gap}}$, can be considered significantly larger than $k_{\text{B}}T$ as well as the energy difference $\Delta E$ of 
%the system when one 
an electron that transfers between the VB and the XC. The latter energy should be small and account for the predicted spectral vicinity of the VB top and the excitonic ground state (see Section~\ref{sec:intro}).

The rates of exciton formation and annihilation are then described by
\begin{equation}
    \gamma^{\text{form}} = C_{\text{form}} \cdot e^{-\kappa}
\end{equation}

\begin{equation}
    \gamma^{\text{anni}} = C_{\text{th}}\cdot e^{-\frac{\Delta E}{k_{\text{B}}T}}\cdot C_{\text{anni}} \cdot e^{-\kappa}
\end{equation}
where $\kappa = \pi\lambda\sqrt{\nicefrac{2m\cdot E_{\text{gap}}}{\hbar^2}}$, the coefficients $C_{\text{form}}$ and $C_{\text{anni}}$ are the tunneling probability amplitudes at the energy $E_{\text{VB}}$, and $C_{\text{th}}$ is the probability amplitude for thermal activation from $E_{\text{XC}}$ that enables tunneling-driven annihilation. Based on this and condensing the unknown coefficients to $A = C_{\text{form}}\cdot e^{-\kappa}$ and $B = C_{\text{th}}C_{\text{anni}}/C_{\text{form}}$, Eq.~\ref{eq:5} becomes

\begin{equation}
\label{effRate}
    \gamma  = A(1-B\cdot e^{-\frac{\Delta E}{k_{\text{B}}T}}).
\end{equation} 
%with $A$, $B$ and $\Delta E$ as explicit parameters. 
%The details on how to obtain this expression can be found in~\ref{AppB}. 

We then use the inverse of Eq.~(\ref{effRate}) to fit the experimental ground state exciton re-formation time shown in \Fref{Fig7}(b) as a function of the transient temperature $T_{\text{t}}$. The resulting fits are shown by the red curves, and agree remarkably well with all data sets at all initial base temperatures. This very good agreement strongly supports the initial assumption that the exciton re-formation rate $\gamma$ extracted from the time-, fluence-, and temperature-dependent data is an effective rate that results from the competition of exciton formation and temperature-dependent annihilation. 

The energy gain in forming a ground state excition, i.e. the $\Delta E$ fit parameter obtained at each initial sample temperature $T_{\text{ini}}$, is reported in Table~\ref{tab1}.
\begin{table}
\begin{center}
\begin{tabular}{|c|c|}
 \hline
 $T_{\text{ini}}$~(K) & $\Delta E$~(eV) \\ [0.5ex] 
 \hline\hline
 110 & 4.35 $\times$10$^{\text{-6}}$\\ 
 \hline
 200 & 4.63 $\times$10$^{\text{-6}}$\\
 \hline
 300 & 4.67 $\times$10$^{\text{-6}}$\\
 \hline
\end{tabular}
\caption{The energy gain in forming a ground state excition, $\Delta E$, obtained from the two quantum-well model fit of the exciton re-formation rate.}
\label{tab1}
\end{center}
\end{table}
We note that for all initial base temperatures $T_{\text{ini}}$, $\Delta E$ is on the order of $10^{-6}$~eV, in agreement with the expectation of energy vicinity of the single-particle VB top and the XC. As outlined in section \ref{sec:intro}, the flattening of the top VB is a result of spontaneous exciton formation leading to an exciton condensate that will continue until the highest-energy electrons are isoenergetic with the XC. The quasi-stationary state, which constitutes the ground state of an EI, is a result of ongoing exciton formation and annihilation at the respective sample temperature. For completeness, we would like to add that with increasing $T_{\text{ini}}$, $\Delta E$ increases. Given the simplicity of the model, that faces complex physics, we refrain from a deeper interpretation of this effect, but note that this behaviour could result from the temperature-induced shrinking of the band gap towards \TC.
%This behavior can be consistently explained by the temperature-induced shrinking of the band gap towards \TC\,leading to upward shifting of the VB top. As a result, the energy splitting between the single-particle VB top and the XC condensate increases for higher $T_{\text{t}}$ and the probability of backward electron tunneling reduces.%, in agreement with the experimental results. 

\section{Summary and Conclusion}
\label{sec:summ}
By means of time- and angle-resolved, one- and two-photoelectron spectroscopy we addressed the occupied and unoccupied electronic states of \TNS, and achieved a full reconstruction of the ultrafast charge carrier and exciton relaxation dynamics photoinduced at temperatures below \TC, where \TNS\,is believed to host an EI phase. By complementary investigation of the photoexcited population of the occupied and unoccupied electronic band structure and a coupled rate equation model, we succeeded to understand how quasiparticles relax in \TNS: photoexcitation initially promotes electrons, originating both from the VB and, even more strongly, from the XC, into the different CBs.  The unoccupied electronic band structure is measured up to a few eV above \EF, a region that was not experimentally explored until now. By that, we showed the occurrence of a saturation of the electron population resonantly excited to high-energy states at $\Gamma$ in corroboration with our previously reported optical absorption saturation by valence electrons at the VB top. The photoexcited electron population is found to decay within less than ca.~100~fs through intraband scattering processes that are likely favored by the downwards dispersion of the transiently occupied CB. The electron population excited at lower energies in the quasi-linearly dispersing CB manifold relaxes on a sub-picosecond time scale, and the decay time follows an inverse linear dependence on the excess energy towards the CB minimum. We discussed this nontrivial behavior as a result of the reduced screening of the Coulomb interaction and the low dimensionality of \TNS. We showed that the across-band-gap relaxation of the photoexcited electrons on a fluence-dependent 0.5 to 2~ps timescale is followed by exciton re-formation. This process is found to compete with the dissociation of ground state excitons caused by the high transient temperatures after photoexcitation. Due to this competition, the effective exciton re-formation time shows a complex dependence on fluence and temperature from few ps up to several 10s of ps. This intricate interplay was described by a two quantum-well model that, despite its simplicity, reproduced the complex data sets with only three free parameters, and thereby confirms the fragile balance of exciton formation and annihilation constituting the EI ground state. Furthermore, it enabled the measurement of the tiny energy barrier on the order of few times $10^{-6}$~eV, which is not only noteworthy from an technical perspective, but also, to our knowledge, the first direct experimental unravelling of quasi-free and exciton-bound electrons in an EI. With this, the presented study resolves the long-standing effort of spectral and temporal disentanglement of the photoinduced dynamics of an XC from the entwined signature of the single-particle VB structure. This progress will open new experimental and theoretical routes to address how many-body interactions
%, such as the weakly-screened electron-hole Coulomb attraction, 
and a subtle energy balance between distinct quasiparticles
%, like quasi-free charge carriers and ground-state excitons,
impact on the ultrafast nonequilibrium behavior of a condensed matter phase.
%. Particularly, the presented findings will hopefully trigger studies on other phases of matter, aiming at disentangling coupled quasiparticle dynamics on the ultrafast time scale. 

\section*{Acknowledgments}
We acknowledge Hide Takagi, Naoyuki Katayama, Minoru Nohara for providing us the single crystals of TNS.
S.M. acknowledges partial financial support through the grant "Finanziamenti ponte per bandi esterni" from Università Cattolica del Sacro Cuore. C.M. acknowledges support from the Swiss National Science Foundation (SNSF) Grant No. P00P\_170597.

%% The Appendices part is started with the command \appendix
%% appendix sections are then done as normal sections
\appendix
\section{Calculation of the pump-induced  temperature increase}
\label{AppA}
The specific heat $C_p$ as a function of temperature is taken from~\cite{Lu17} and fitted with a 10$^{\text{th}}$ order polynomial. Now, since $C_p = \nicefrac{dQ}{dT}$, one can rewrite it $C_p~dT = dQ$ and integrate this equation: 
\begin{equation}
    Q = \int_{T_{\text{ini}}}^{T}C_{p}(t)dt
\end{equation}
This is an integral equation for the temperature $T$ that is eventually reached when the pulse deposits the heat $Q$, starting from initial basis temperature $T_{\text{ini}}$. By solving it, we obtain the increase in temperature, for different initial base temperatures and pump fluence values. The deposited heat is obtained from the surface-transmitted fluence using an absorption length of 100~nm estimated from values of optical permittivity and conductivity given in~\cite{Lar17}.

%\section{Two-quantum-well model}
%left out for now
%%

%\bibliography{Publications-2021_TNS_4PSS.bib}
%\bibliographystyle{elsarticle-num} 
%% else use the following coding to input the bibitems directly in the
%% TeX file.

%\begin{thebibliography}{00}

% %% \bibitem{label}
% %% Text of bibliographic item

% \bibitem{}

%\end{thebibliography}

\end{document}